\documentclass[twocolumn,prd,showpacs,10pt]{revtex4}
\usepackage{graphicx}% Include figure files
\usepackage{dcolumn}% Align table columns on decimal point
\usepackage{bm}% bold math

\def\gsim{\stackrel{>}{\sim}}
\def\wm{w_Q^{m}}
\def\w0{w_Q^{0}}
\def\at{a_c^{m}}
\def\s8{\sigma_8}

\begin{document}

\title{Model-independent dark energy test with $\sigma_8$ using 
results from the Wilkinson Microwave Anisotropy Probe}

\author{Martin Kunz${}^1$, Pier-Stefano Corasaniti${}^{2,3}$, David Parkinson${}^4$ and Edmund J. Copeland${}^2$}
\address{${}^1$ Astronomy Centre, University of Sussex,
  Brighton, BN1 9QJ, UK}
\address{${}^2$ Department of Physics and Astronomy, University of Sussex,
  Brighton, BN1 9QJ, UK}
\address{${}^2$ ISCAP, Columbia University, Mailcode 5247,
  New York NY 10027, USA}
\address{${}^4$ Institute of Cosmology and Gravitation, University of
  Portsmouth, Portsmouth, PO1 2EG, UK}

\begin{abstract}
By combining the recent WMAP measurements of the cosmic microwave background
anisotropies and the results of the recent luminosity distance measurements to
type-Ia supernovae, we find that the normalization of the matter power spectrum on 
cluster scales, $\sigma_8$, can be used to discriminate between
dynamical models of dark energy (quintessence models) and a 
conventional cosmological constant model ($\Lambda$CDM). 
\end{abstract}

\maketitle

\section{Introduction} 
The WMAP satellite measurements of the Cosmic Microwave Background
anisotropies \cite{BENNETT} have 
provided accurate determinations of many of the fundamental cosmological parameters. 
When 
combined with other data sets such as the luminosity distance to type-Ia 
supernovae 
or large scale structure (LSS) data \cite{PEL,TONRY,KNOP,ef},
they reinforce the need for an exotic form of dark energy, which 
is characterized by a negative pressure and is responsible for the observed
accelerated expansion of the universe.
There are two main scenarios used to explain the nature
of the dark energy, a time independent
cosmological constant $\Lambda$ and quintessence, 
which involves an evolving scalar field Q \cite{WETTERICH,ZATLEV,AMS1}. 
Previous tests of quintessence with
pre-WMAP CMB data \cite{CORAS1,MORT,BRUCE}, have 
led to constraints on the value of the dark energy equation of state parameter, 
$w_Q \lesssim -0.7$ 
with the cosmological constant value, $w_\Lambda=-1$ being the best fit.
Nevertheless a dynamical form of dark energy is not excluded.
Specifically the detection of a time variation in this parameter 
would be of immense importance as it would rule
out a simple cosmological constant scenario.
When parameterising quintessence
models we do not want to assume simply a constant equation of state $w_Q$ 
since this introduces a systematic
bias in the analysis of cosmological distance measurements \cite{MAOR},
with the effect of favouring larger negative values of $w_Q$ if the
dark energy is time dependent.  For instance it is possible that
claims for a `phantom' component, where $w_Q < -1$ \cite{MORT,CALDWELL} 
are entirely caused
by this effect. Moreover assuming $w_Q$ constant underestimates
the contribution of the dark energy perturbations (which are a specific feature of scalar
field models) on the evolution of
the gravitational potentials and consequently the effect on the CMB
power spectrum \cite{CORAS3}. 
In this paper we deliberately do not assume $w_Q$ to be constant, 
rather we focus on the relation between a dynamical 
dark energy component and
the normalisation of the dark matter power spectrum on cluster scales,
$\sigma_8$. We also discuss the age of the universe, $t_0$, and
show how the new data sets undermine its use for distinguishing between
different dark energy models.

\section{Method and data} 
In this analysis, rather than considering a specific scalar field model, 
we allow for a time dependence of the equation of state parameter $w_Q$. 
Several formulae have been proposed in the literature \cite{GERKE,LINDER}
all with limited applicability. In \cite{CORAS2} 
a form for $w_Q(z)$ was suggested which is valid at all redshifts and parameterises
the equation of state in terms of five parameters, which specify
the value of the equation of state parameter today
$\w0$, and during the matter/radiation eras $\wm/w_Q^r$; 
the scale factor $\at $ where the equation of state changes from $\wm$ to $\w0$
and the width of the transition $\Delta$.
Since Big-Bang Nucleosynthesis bounds limit the amount of dark energy to be
negligible during the radiation dominated era, without loss of generality
we can further reduce our parameter space by setting $w_Q^r=\wm$ 
in Eq.~(4) of Ref.~\cite{CORAS2}.
The parameters given
by the vector $\overline{W}_{Q} = (\w0, \wm, \at, \Delta)$ 
can account for
most of the dark energy models proposed in the literature. For instance
quintessence models characterized by a slowly varying equation of state, such 
as supergravity inspired
models \cite{BRAX}, correspond to a region of our parameter space for which
$0<\at/\Delta<1$, while rapidly varying models, such as 
the two exponential potential case \cite{NELSON},
correspond to $\at/\Delta>1$. Models with a simple constant equation of state
are given by $\w0=\wm$. The cosmological constant case is also included 
and corresponds to the following cases: $\w0=\wm=-1$ or $\w0=-1$ and 
$\at\lesssim 0.1$ with $\at/\Delta>1$.
Assuming a flat geometry we perform a likelihood analysis of the WMAP data to constrain
dark energy models specified by the vector $\overline{W}_{Q}$ and the
cosmological parameters $\overline{W}_{C}=(\Omega_Q,\Omega_b h^2,h,n_S,\tau,A_s)$ which
are the dark energy density, the baryon density, the Hubble parameter, the scalar
spectral index, the optical depth and the overall amplitude of the scalar fluctuations
respectively. We have modified a version of the 
CMBFAST code \cite{ZALDA} to include the dark energy perturbation equations
in terms of the time derivatives of the equation of state \cite{BOB}.
In order to break the geometric degeneracy between $\w0$, $\Omega_Q$ and $h$,
we use the most recent compilation of supernova data of \cite{TONRY} in addition 
to the WMAP TT and TE power spectrum data. We evaluate the likelihood of
CMB data with the help of the software provided by the WMAP team \cite{VERDE}.
The important point which we want to stress is that we are able to treat
both data sets, (WMAP and SN-Ia) 
without making any prior
assumptions concerning the underlying
cosmological model, in
order to be as conservative as possible and to evade potential problems
with issues like relative normalisations and bias.
We restrict our analysis to dark energy models that satisfy the null
dominant energy condition and
$\w0,\wm \geq -1$ and
following the analysis by the WMAP team, we use
the prior $\tau \leq 0.3$ in order to prevent $\Omega_b$ from taking
unphysically high values.

\section{Results} 
\begin{figure}[htb]
\begin{center}
\includegraphics[width=80mm]{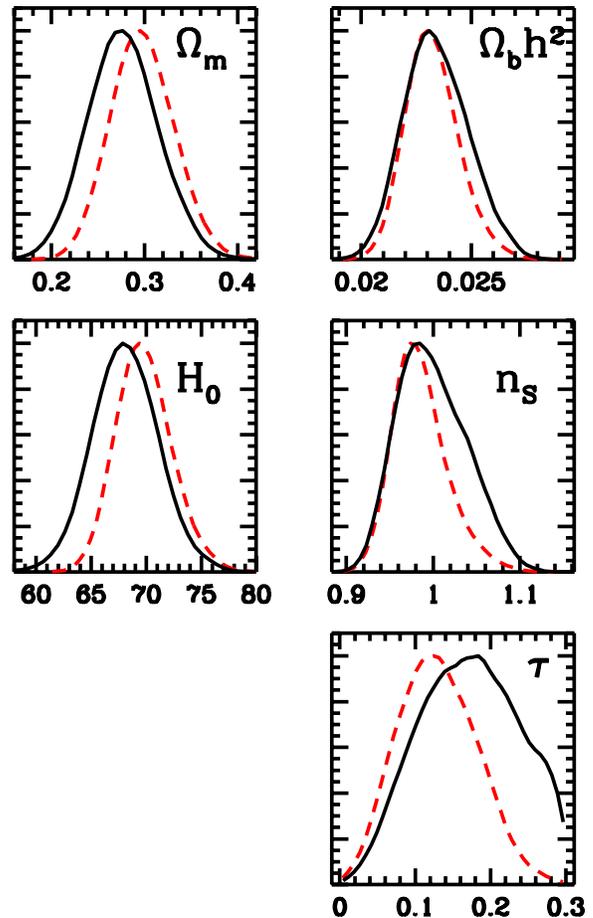} \\
\caption[cpar]{\label{cpar} Marginalized likelihoods for the various
cosmological parameters in the $\Lambda$CDM scenario 
(red dashed curve) and including
the QCDM models (black curve). The results agree very well
with each other.}
\end{center}
\end{figure}

The WMAP CMB data constrains the cosmological parameters
$\overline{W}_{C}$ in a range of values consistent with
the results of previous analysis such as \cite{SPERGEL,LUCA,DORAN}.
In particular we find the scalar spectral index $n_S=1.00\pm0.04$,
the physical baryon density $\Omega_b h^2=0.0234\pm0.0014$ and
the optical depth $\tau=0.17\pm0.06$.
As mentioned above, in order to break the degeneracy between 
$\w0$, $\Omega_Q$ and $h$, we combine the CMB data with the SN-Ia luminosity
distance measurements. This allows us to constrain the Hubble constant to be
$h=0.68\pm0.03$, in 
agreement with the HST value \cite{FREEDMAN}, the dark energy density
$\Omega_Q=0.72\pm0.04$ (all limits so far at 1$\sigma$) and the
present value of the equation of state $\w0<-0.82$ (at 95\% CL). It is important to stress 
that the the addition of the dark
energy parameters $\overline{W}_{Q}$ does not introduce 
any new degeneracies with the other 
parameters. This is clear from the fact that the constraints on 
$\overline{W}_{C}$ are in agreement with other previous data analyses.
Figure \ref{cpar} shows the marginalised one-dimensional likelihoods
for $\Lambda$CDM and the dynamic dark energy models.
We will defer a detailed discussion of these results to a later
paper, and in this paper concentrate 
on the use of dark matter clustering as a probe
of quintessence models. 

\begin{figure}[h]
\begin{center}
\includegraphics[width=80mm]{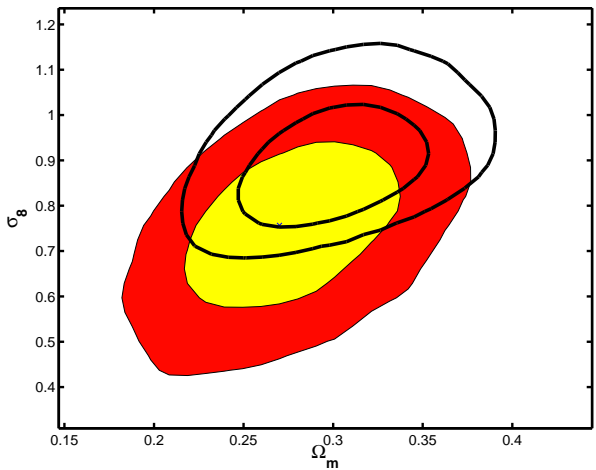} \\
\caption[oms8]{\label{oms8} Marginalized 68\% and
95\% confidence contours for quintessence (filled contours) and
$\Lambda$CDM models (solid lines). $\Lambda$CDM
has a systematically higher value of $\s8$, and
a slightly higher value of $\Omega_m$.}
\end{center}
\end{figure}

In general we expect dark energy to affect the
value of $\sigma_8$ because it can
lead to a different expansion history of the universe \cite{doransigma}.
However, in \cite{CORAS3} it was shown that different
dark energy models leave particular imprints on the large
angular scales of the CMB anisotropy power
spectrum through the integrated Sachs-Wolfe (ISW) effect. 
The excess of power produced by the ISW at  low multipoles affects
the normalization of the matter power spectrum \cite{bartel}. 
For instance models
with a fast late time transition in the equation of state produce a larger
ISW effect than a pure cosmological constant scenario. As a consequence
they require a smaller amplitude of  primordial fluctuations
in order to match the observed CMB spectrum. In this case the
predicted value of $\sigma_8$ will be smaller than in the $\Lambda$CDM model.
This specific class of models has already been investigated using pre-WMAP data 
\cite{BRUCE,CONDENSATION}, but the results underestimated 
the optical depth subsequently found by WMAP, leading to an over-estimation 
of the power on small angular scales.
It is only with the release of
the first year of WMAP data that through one CMB data set, we can link the
anisotropies on large and small angular scales. 
This is an exciting feature of the data, 
as it allows us to properly assess the effects of ISW and the normalization
of the matter power spectrum.
In figure \ref{oms8} we plot the two dimensional likelihood 
contours in the $\Omega_m-\sigma_8$ plane.
The filled
contours correspond to $1$ and $2\sigma$ values for the dark energy models spanned by
$\overline{W}_{Q}$, while the solid curves correspond to the $\Lambda$CDM case.
As expected from the above discussion, we 
note that $\Lambda$ models have systematically higher values of
$\sigma_8$ than models with  a time varying equation of state.

\begin{figure}[h]
\begin{center}
\includegraphics[width=80mm]{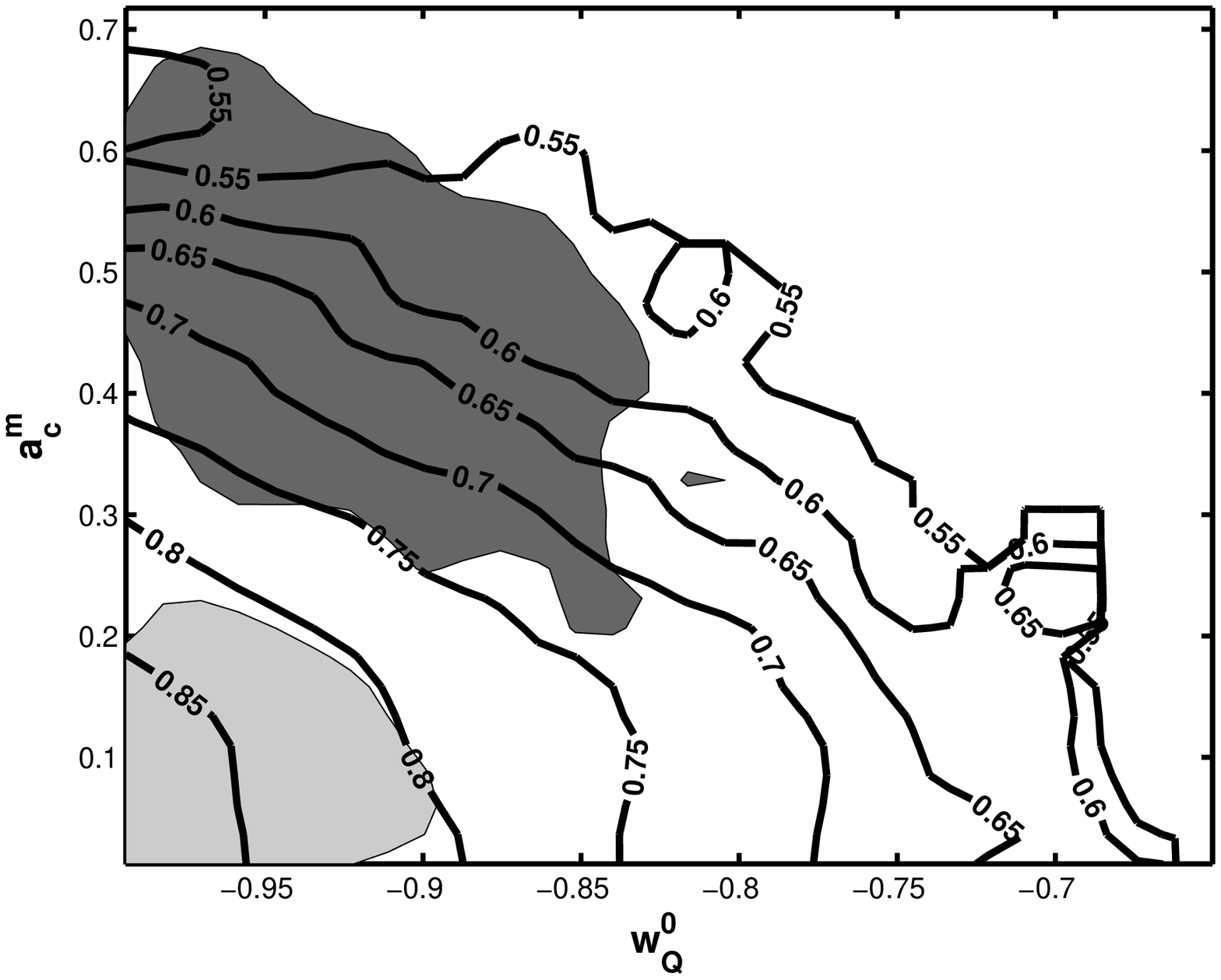} \\
\caption[atw0]{\label{atw0} The average $\s8$ as a function
of $\w0$ and $\at$ for models with a rapid transition in
$w_Q$ (numbered lines). We also show the 68\% confidence
regions for models with $\s8 < 0.6$ (dark gray) and $\s8 > 0.9$
(light gray).}
\end{center}
\end{figure}

\begin{figure}[h]
\begin{center}
\includegraphics[width=80mm]{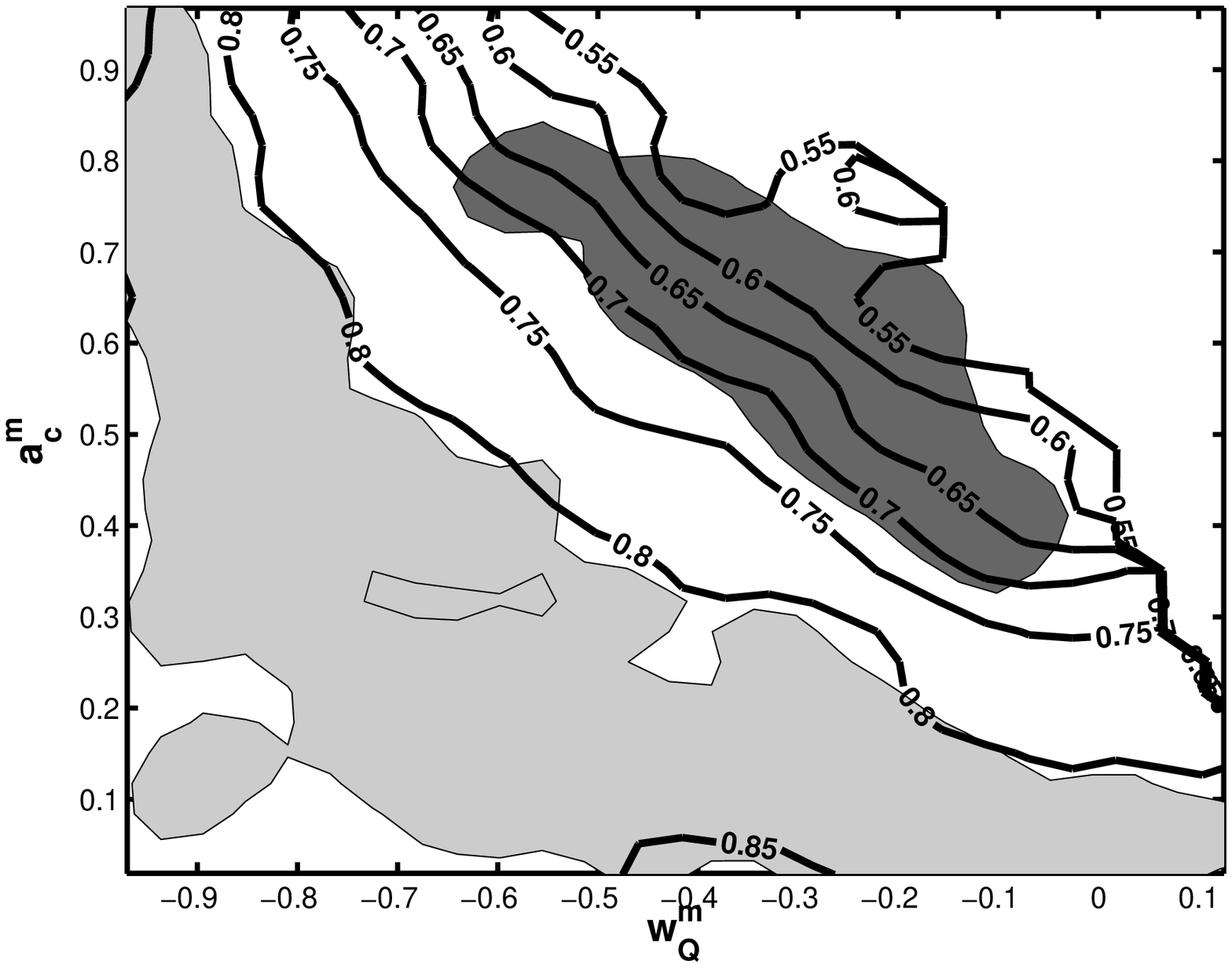} \\
\caption[wm]{\label{wmac} The average $\s8$ as a function
of $\wm$ and $\at$ for models with a rapid transition in
$w_Q$ (numbered lines). We also show the 68\% confidence
regions for models with $\s8 < 0.6$ (dark gray) and $\s8 > 0.9$
(light gray).}
\end{center}
\end{figure}

\begin{figure}[h]
\begin{center}
\includegraphics[width=80mm]{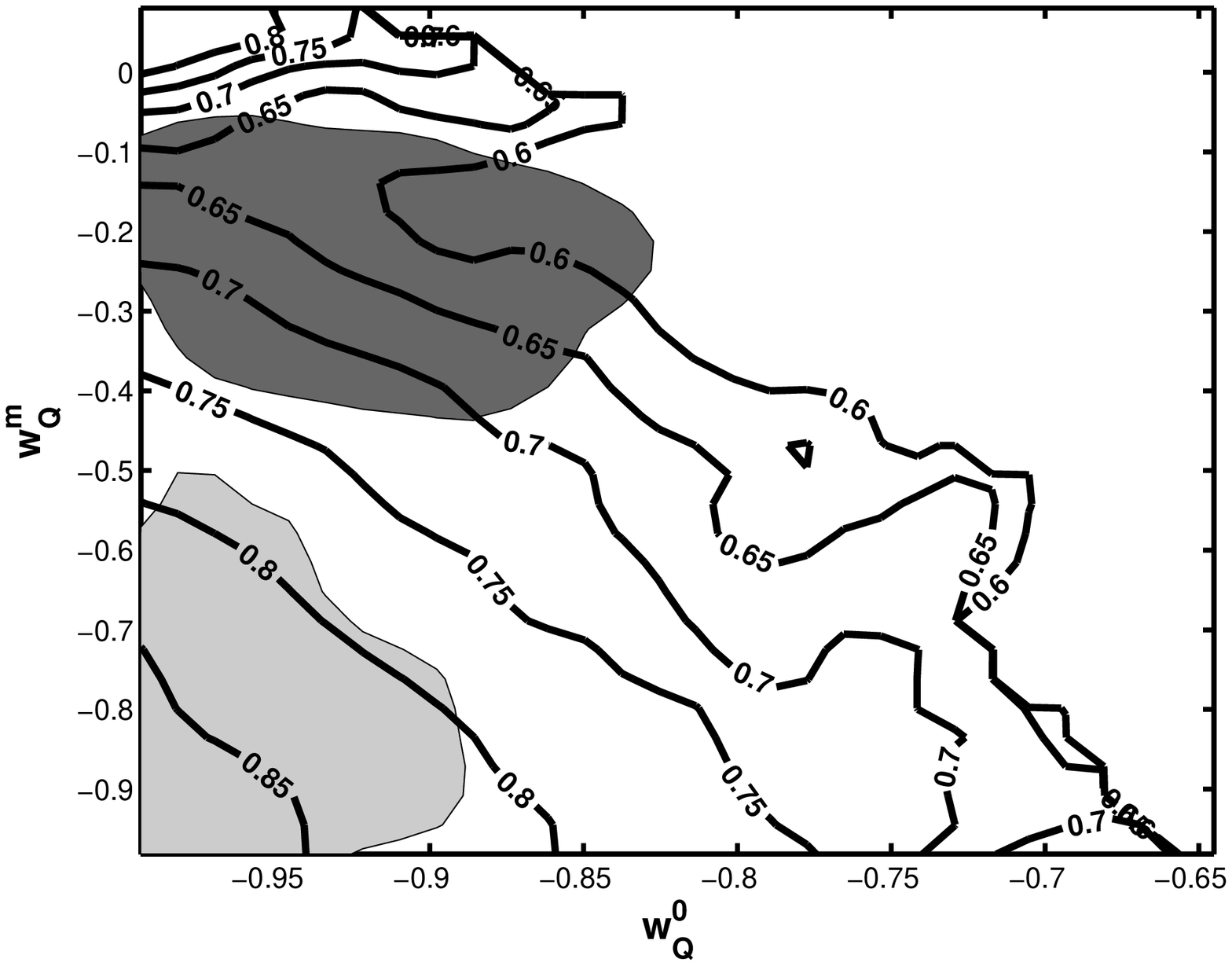} \\
\caption[wmw0]{\label{wmw0} The average $\s8$ as a function
of $\w0$ and $\wm$ for models with a smoothly varying
$w_Q$ (numbered lines). We also show the 68\% confidence
regions for models with $\s8 < 0.6$ (dark gray) and $\s8 > 0.9$
(light gray).}
\end{center}
\end{figure}

It seems clear that a CMB independent estimate of the value of $\sigma_8$ would be
able to distinguish between a $\Lambda$CDM and dynamical equation of state  model. For instance values of $\sigma_8<0.7$ would be rejected at 2$\sigma$ in the $\Lambda$CDM case. 
More specifically we find that the value of $\sigma_8$ can discriminate
between different dark energy models. This can be seen in
figures \ref{atw0}, \ref{wmac} and \ref{wmw0} which are the main result
of this paper. 
In Fig.~\ref{atw0} we plot the average value of $\sigma_8$
as a function of $\at$ and $\w0$, where the average is taken
over all models in our chain which exhibit a rapid transition
(defined here as $\wm>-0.2$ and $\Delta<0.1$). A  $\Lambda$CDM 
model corresponds to $\at\rightarrow0$ and $\w0=-1$. The average
value of $\sigma_8$ in this point is $0.9$. As we move away
from the $\Lambda$CDM corner, the average $\sigma_8$ decreases
monotonically, as seen by the contours. 
To assess the usefulness of $\s8$ for
distinguishing between models given todays data, we also
plot two 68\% confidence regions, one for models with
$\s8 > 0.9$ (lighter gray) and one with $\s8 < 0.6$ (darker gray).
Clearly, if we
restrict ourselves to models with a high value of $\s8$, we
favour a $\Lambda$CDM-like behaviour of the dark energy.
In the opposite case, we find $\at\gsim0.3$. Together with
the fast-transition conditions given above, this means that
these models have an equation of state $w(z>2)\gg -1$,
and we would exclude the case $p=-\rho$ at over 95\% CL. 
As we marginalise
over all other parameters, we see that no degeneracies
spoil this result.

As a complementary view, we can plot $\at$ and $\wm$ for
fast-transition models (without the condition on $\wm$),
see Fig.~\ref{wmac}.
The data requires that $\w0 < -0.8$ and so $\Lambda$CDM
models occupy the region defined by either $\at \rightarrow 0$
(in which case the equation of state is independent
of $\wm$)
or $\wm \rightarrow -1$ (and thus $w(z) \approx -1$
without transition), which again coincides with the
high-$\s8$ models. Models with $\s8 < 0.6$ on the other
hand require both $\at \gsim 0.3$ and $\wm\gsim-0.7$ at
68\% CL.

Fig.~\ref{wmw0} is the corresponding figure for
dark energy models with a slowly varying
equation of state ($0<\at/\Delta<0.8$). In this case
the relevant parameters are $\wm$ and $\w0$, and the $\Lambda$CDM 
models are now at $\w0=\wm=-1$. Again, $\s8$ decreases
rapidly as we move away from that corner. We show once more
the $1\sigma$ regions for models with 
$\s8 > 0.9$ (lighter gray) and with $\s8 < 0.6$ (darker gray).
Models with a high value of $\s8$ are again clustered
around the $\Lambda$CDM region, and those with a low clustering
amplitude require $w \gg -1$ at high redshift.

We expect these regions to shrink as the cosmological
parameters become more constrained by future data, 
which will improve the impact of clustering as a
probe of the time dependence of the dark energy.
This is our main result, and it means that, given a
precise measurement of $\s8$, we can impose strong
limits not only on the value of $w$ today, {\em but
also at earlier times}. Even if $\w0\approx -1$ today,
we are able to probe its behaviour at higher redshift
and to either exclude $\Lambda$CDM or significantly
constrain quintessence type models. Although especially
slowly varying models cannot be ruled out as they can
approximate the behaviour of a true cosmological constant
arbitrarily closely, these models become less and less
attractive as they start to require the same fine tuning
as $\Lambda$ itself.
 
Why are we using $\sigma_8$ as a variable as opposed to simply choosing one 
of the many published measured values of $\sigma_8$? 
First, the published data shows a large spread of values \cite{BRIDLE}, 
so that our conclusions would
strongly depend on the choice of data sets.
Secondly, the measurements also depend in general on the
dark energy parameters and the results quoted are only
valid for $\Lambda$CDM models. For example, this is the case  for 
the large scale structure results,
which implicitly assume a $\Lambda$CDM model when passing from redshift
space to real space, and for weak lensing measurements. In the second
case, the dependence on the dark energy characteristics is strong
enough that it can be used to constrain the evolution of the equation of
state \cite{BENABED}.
As an illustration, we can assume that the clustering results
deduced from velocity fields in Ref.~\cite{FELDMAN} are unaffected by the
details of the dark energy evolution. As a rough approximation to
their PSCz results, we set $\s8 \approx (1.13\pm0.05) (\Omega_m/0.3)^{0.6}$.
In this case, the constraints on quintessence models become much
stronger, e.g.~$\w0 < -0.9$ at 95\% CL.
On the other hand, if future precision measurements converge on
$\s8 \lesssim 0.7$ then $\Lambda$CDM is ruled out at high significance.

Moreover, $\s8$ is linked to the
amplitude of the matter power spectrum $P(k)$ on small scales.
To measure 
a possible running of the scalar spectral index, $dn_S/d\log k$,
in inflationary models,
it is necessary to combine CMB data on large scales with $P(k)$
on small scales.
Since quintessence models can change the amount of clustering on
small scales with respect to a $\Lambda$CDM model, it
is possible for them to mimic the effect of such a running.
This possibility should be kept in mind when constraining models
through the combination of different data sets \cite{doransigma}.

\begin{figure}[h]
\begin{center}
\includegraphics[width=80mm]{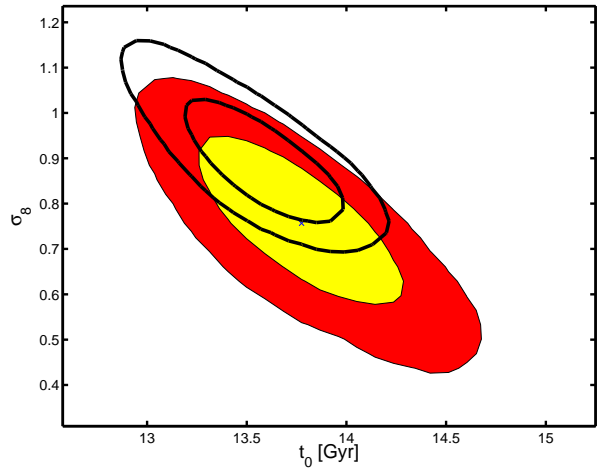} \\
\caption[t0s8]{\label{t0s8} Marginalized 68\% and
95\% confidence contours for quintessence (filled contours) and
$\Lambda$CDM models (solid lines).}
\end{center}
\end{figure}

Another observable which has been studied in this context
is the age of the universe, $t_0$ \cite{KRAUSS,JIMENEZ}, which is
in general also a function of the dark energy parameters 
$\overline{W}_{Q}$. An independent measurement of $t_0$ 
(for which the WMAP limit does {\em not} qualify, as it
explicitly assumes $\Lambda$CDM) can thus be used to set
limits on the equation of state. Since the luminosity
distance $d_L$ and $t_0$
possess a similar dependence on the Hubble rate, the SN-Ia data,
which probe about two-thirds of the age of universe,
can provide tight constraints on  $t_0$ even for generic dark energy models.
For example in \cite{TONRY} considering  
$\Lambda$CDM cosmologies, the authors obtain $H_0 t_0 = 0.96\pm0.04$.
The limit
is also valid for quintessence, as we find $H_0 t_0 = 0.96\pm0.03$ for the
combination of CMB and SN-Ia data. This constraint, together
with the remaining slight degeneracy in $H_0$ which leads
to lower values of the Hubble constant as we move away from
the $\Lambda$CDM models, means that the allowed quintessence
models are older than the those with a cosmological constant, as
we can see in Fig.~\ref{t0s8}. The marginalised age of quintessence 
universes is $t_0=13.8\pm0.3$ Gyr, while in the
$\Lambda$CDM case $t_0=13.55\pm0.26$. Clearly, it will
be difficult to use $t_0$ to disentangle different models until
the uncertainty in the cosmological parameters is further reduced.
But if we were to find a lower limit on the age of
the universe which is too high for $\Lambda$CDM, we could potentially
interpret it to be a sign of quintessence.

\section{Conclusions}
In this paper we have demonstrated how, by combining
WMAP and SN-Ia data, 
it is possible to use the normalisation of the dark energy
 power spectrum on cluster scales, $\sigma_8$, to discriminate between
dynamical models of dark energy (quintessence models) and a conventional 
cosmological constant model ($\Lambda$CDM). 
In particular we have shown 
for the first time that a CMB independent measurement of $\sigma_8$
allows us to constrain the
parameters describing the evolution of the dark energy equation of state.
For instance, we found that standard $\Lambda$CDM is ruled out at over
95\% CL (compared to a time dependent dark energy component)
if $\sigma_8<0.7$. This constraint can be relaxed by going beyond
the standard model, i.e. introducing very massive neutrinos or a
running of the spectral index \cite{ALLEN}. However, we expect
improved data to lead
to stronger limits in the near future. We have also briefly 
discussed the use of the age of the universe $t_0$ as a way of 
constraining dark energy models, and shown that by itself it does not
discriminate between quintessence and $\Lambda$CDM models, 
although coupled with $\sigma_8$, it may act as a useful cross check.

\begin{acknowledgments}
We thank R.R.~Caldwell, M.~Doran and K.~Moodley for useful
discussions. MK and DP are supported by PPARC, 
PSC was partially supported by a Sussex University bursary.
We acknowledge extensive use of the UK National Cosmology Supercomputer funded
by PPARC, HEFCE and Silicon Graphics / Cray Research.

\end{acknowledgments}

\end{document}